# *Topological Angular Momentum and Radiative Heat Transport in Closed Orbits*

*Mário G. Silveirinha*[*]

[(1)]*University of Lisbon–Instituto Superior Técnico and Instituto de Telecomunicações, Avenida Rovisco Pais, 1, 1049-001 Lisboa, Portugal*

**Abstract**

Here, we study the role of topological edge states of light in the transport of thermally generated radiation in a closed cavity at a thermodynamic equilibrium. It is shown that even in the zero temperature limit – when the field fluctuations are purely quantum mechanical – there is a persistent flow of electromagnetic momentum in the cavity in closed orbits, deeply rooted in the emergence of spatially separated unidirectional edge state channels. It is highlighted the electromagnetic orbital angular momentum of the system is nontrivial, and that the energy circulation is towards the same direction as that determined by incomplete cyclotron orbits near the cavity walls. Our findings open new inroads in topological photonics and suggest that topological states of light can determine novel paradigms in the context of radiative heat transport.

---

[*] To whom correspondence should be addressed: E-mail: *mario.silveirinha@co.it.pt*

# I. Introduction

The discovery of topological states of light uncovered a myriad of physical platforms wherein the wave propagation is impervious to perturbations of the propagation path and immune to back-scattering [1-17]. In particular, Chern-type photonic insulators are characterized by a non-trivial topological number – the Chern invariant – which being an integer is absolutely insensitive to continuous deformations of the material response, e.g., to weak continuous changes in the microscopic or nanoscopic constituents [1-4, 18-20]. The Chern number can only vary when some topological charge is exchanged between different photonic bands, i.e., when a bandgap closes and reopens [1-3]. Thus, a topological phase transition must occur when some material is continuously transformed into another material with a different Chern number (e.g., through a continuous deformation of the structural unities or a variation of a biasing field). Notably, this property implies that if two topologically distinct materials that share a common bandgap are joined to form an interface, then topologically protected *unidirectional* edge states will appear in the bandgap, a property which is known as the "bulk edge correspondence principle" [3, 20, 21].

A nonzero Chern number requires breaking the time-reversal symmetry and thus a nonreciprocal material response. This can be achieved using a static magnetic field that originates a gyrotropic response [1, 7]. Nonreciprocal effects may also occur in systems with moving parts [17, 22, 23], when the material response is time-modulated [24], or in presence of strong nonlinearities [25]. Furthermore, in some antiferromagnetic materials the time-reversal symmetry may be spontaneously broken due to magnetic ordering [18, 26].

It is natural to wonder whether the intriguing properties of topologically protected unidirectional edge states can have any exotic consequences in the context of fluctuational electrodynamics, i.e., when the electromagnetic fields are either generated by the random jiggling of atoms due to thermal agitation or due to purely quantum fluctuations. Specifically, we are interested in scenarios where the system of interest is closed (e.g., a closed cavity) and is in a thermal equilibrium. The presence of topologically protected "one-way" channels raises puzzling questions: do these channels lead to a transport of the thermal (or zero-point) energy? Does a heat transport prevent that a thermal equilibrium is reached or does it breach the second law of thermodynamics due to the energy concentration in some location of the cavity? Here, it is theoretically shown that the emergence of topological edge states *does not* lead to any paradoxical settings and that standard fluctuation electrodynamics provides a conclusive answer to the enunciated questions. Specifically, it is found that the topological edge states induce a circulation (with no net sinks or sources) of thermal (or zero-point) energy in closed orbits, such that the *angular momentum* of the electromagnetic field is nonzero. Remarkably, it is shown that the heat flux circulates towards the direction determined by incomplete electron cyclotron orbits. Notably, the energy flow persists even in the zero temperature limit, wherein the system is in its (parametric) "ground" state.

It is important to connect our findings with previous studies. The circulation of electromagnetic momentum in nonreciprocal systems in a thermal equilibrium has been discussed in a few different contexts. In Ref. [27], we demonstrated that in a system with parts in a shear motion the quantum expectation of the zero-point electromagnetic momentum associated with a moving body is typically nonzero, even though the expectation of the total momentum of the system vanishes. Unfortunately, in such systems the effect is negligibly small for nonrelativistic velocities. More recently [28], it was shown that in a nonreciprocal many-body system formed by magnetized nanoparticles near-field interactions enable a persistent directional heat current at a thermodynamic equilibrium, in qualitative agreement with our conclusions. This article builds on these previous works and looks at the problem from a different angle: we investigate the heat circulation in a system with topological properties, showing that the topological unidirectional edge states lead to the formation of spatially separated channels wherein the energy flows in closed orbits, so that the angular electromagnetic momentum is nontrivial at equilibrium.

It is relevant to mention that the influence of a nonreciprocal electromagnetic response in the context of radiative heat transport in nonequilibrium situations (e.g., in presence of temperature gradient) was discussed in other recent works. For example, it was shown that nonreciprocal effects enable a giant radiative thermal rectification [29], a thermal Hall effect [30] and a tunable near-field heat transfer [31]. Asymmetric radiative heat flows can also occur in reciprocal structures with a temperature dependent response [32-33] or in anisotropic materials [34]. More generally, there is a great interest in the recent literature in novel approaches to tame the radiative heat flow at the nanoscale due to its applications in thermophotovoltaics [35-41],thermal radiation based microscopy [42-46] or radiative cooling [47].

## II. Gyrotropic Waveguide

We consider a closed cavity formed by a parallel plate waveguide filled with an electric gyrotropic material (Fig. 1a). The widths of the structure along the *x* and *y* directions are much larger than the waveguide height ($L_x, L_y \gg d$) and, for simplicity of modeling, the corresponding ends are terminated with periodic boundary conditions. Thus, the structure can be assumed invariant to translations along the *x* and *y* directions. The waveguide walls are modeled as perfect electric conductors (PEC), which is a good approximation up to terahertz frequencies.

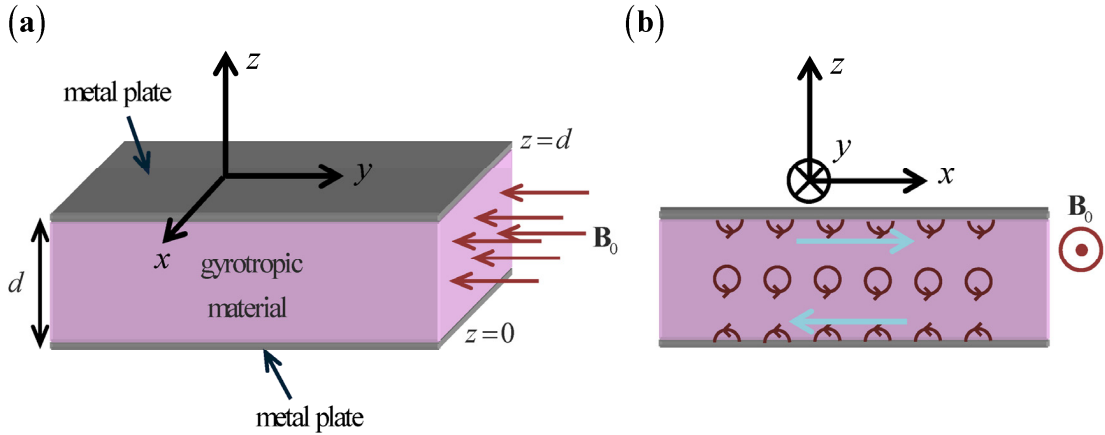

Fig. 1. (Color online) **Geometry of the parallel-plate gyrotropic waveguide.** (a) The waveguide is filled with a gyrotropic material biased with a uniform static magnetic field oriented along the negative *y*-direction. (b) Sketch of the electron cyclotron orbits in the gyrotropic material. The orbits are incomplete at the metallic plates inducing and electron flow to the right (left) on the top (bottom) plate.

The gyrotropic material is biased with a static magnetic field ($\mathbf{B}_0 = -B_0 \hat{\mathbf{y}}$) directed along the negative *y*-direction, so that the relative permittivity tensor is of the form $\overline{\varepsilon} = \varepsilon_t \mathbf{1}_t + \varepsilon_a \hat{\mathbf{y}}\hat{\mathbf{y}} + i\varepsilon_g \hat{\mathbf{y}} \times \mathbf{1}$ with $\mathbf{1}_t = \mathbf{1} - \hat{\mathbf{y}}\hat{\mathbf{y}}$. It is assumed that the frequency dispersion of the permittivity elements is the same as in a magnetized electric plasma (e.g., InSb) with negligible material loss:

$$\varepsilon_t = 1 - \frac{\omega_p^2}{\omega^2 - \omega_0^2}, \quad \varepsilon_a = 1 - \frac{\omega_p^2}{\omega^2}, \quad \text{and} \quad \varepsilon_g = \frac{-1}{\omega}\frac{\omega_p^2 \omega_0}{\omega_0^2 - \omega^2}. \tag{1}$$

Here, $\omega_p$ is the plasma frequency and $\omega_0 = -qB_0/m$ is the cyclotron frequency determined by the biasing field ($q = -e$ is the negative charge of the electrons and $m$ is the mass) [48].

## A. *Topological properties of the bulk gyrotropic material*

Figure 2a shows the band structure of a magnetized plasma with $\omega_0 = 0.5\omega_p$ for different directions of propagation ($\chi = 0º$, $\chi = 45º$ and $\chi = 90º$). The band diagrams were obtained with the formalism described in Appendix A [Eq. (A4)]. The angle $\chi$ is measured with respect to the *y*-direction, i.e., a direction parallel to the bias magnetic field. Note that generally $\chi = const.$ determines a conical surface; for $\chi = 90º$ the conical surface reduces to the *xoz* plane whereas for $\chi = 0º$ it degenerates into the *y*-axis. Thus, Fig. (2a*i*) represents the band structure for propagation in the *xoz* plane. It is well known that when the propagation direction is perpendicular to the bias field the waves are either TE-polarized (with electric field along the *y*-direction) [green dashed line in Fig. (2a*i*)] or TM-polarized (with electric field in the *xoz* plane and magnetic field along the *y*-direction) [blue lines in Fig. (2a*i*)] [19, 20, 49-51]. In particular, it is seen that TM-polarized waves have two different bandgaps for propagation in the *xoz* plane. Here, we will focus on the high-frequency bandgap determined by $\omega_{gL} < \omega < \omega_{gU}$ with $\omega_{gL} = \sqrt{\omega_0^2 + \omega_p^2}$ and $\omega_{gU} = \omega_0/2 + \sqrt{(\omega_0/2)^2 + \omega_p^2}$, delimited by the dashed gray horizontal grid lines in Fig. 2a. For propagation in the *xoz* plane this bandgap separates the TM-polarized waves into two subsets of topologically distinct bands (the TE-band is topologically trivial).

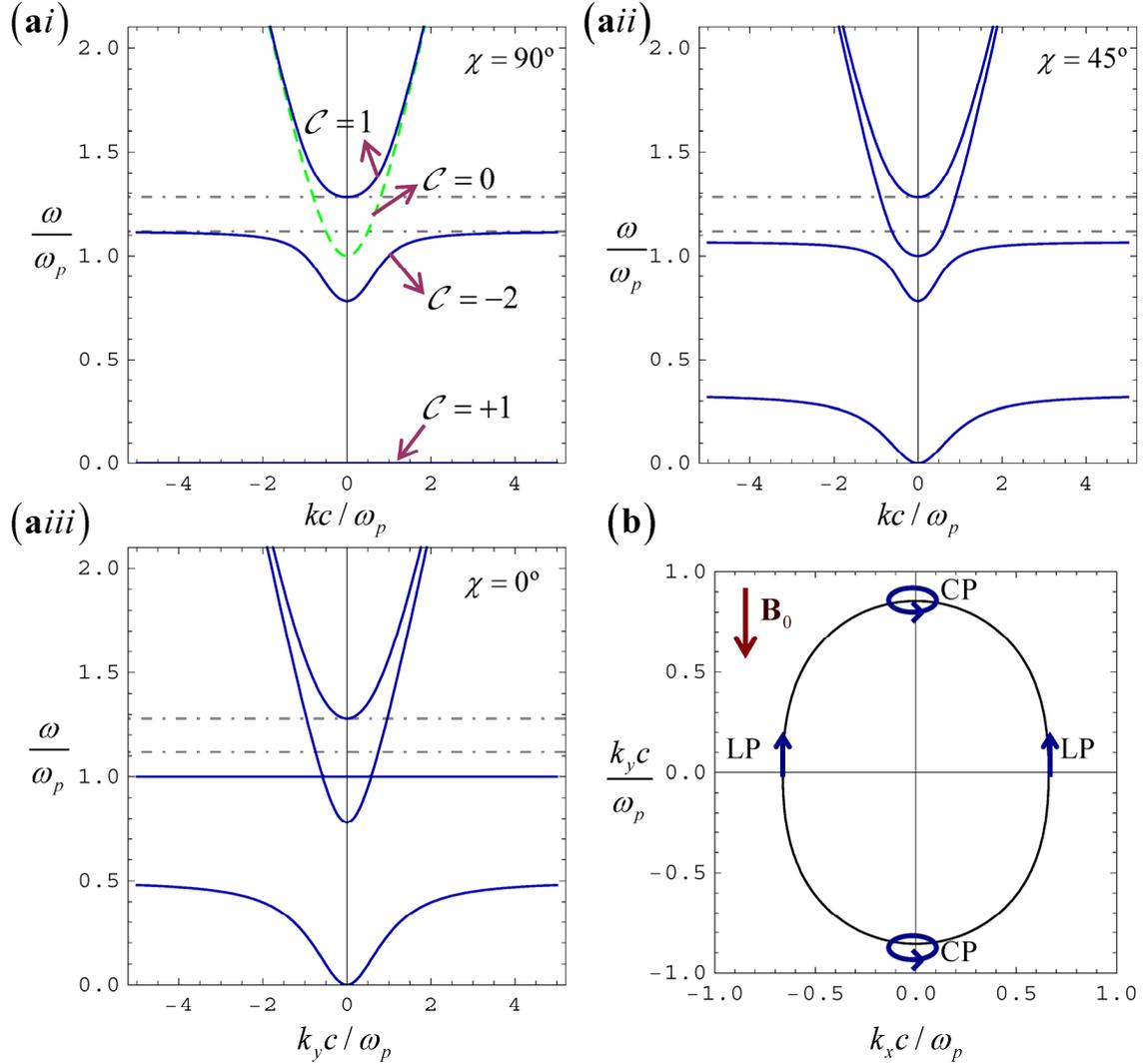

Fig. 2. (Color online) **Band structure and isofrequency contour for a bulk magnetized plasma.** (a) band structure for propagation along the direction (i) $\chi = 90°$, (ii) $\chi = 45°$, and (iii) $\chi = 0°$ measured with respect to the bias magnetic field. In panel (a*i*), the TE-band is depicted with a dashed green line whereas the remaining bands (solid blue lines) are associated with TM-polarized waves; the insets indicate the Chern numbers. (b) Isofrequency contour for $\omega = 1.2\omega_p$ and propagation in the *xoy* plane.

There is a single band above the gap with Chern number $\mathcal{C}_3 = 1$ and two bands below the gap with total Chern number $\mathcal{C}_1 + \mathcal{C}_2 = +1 - 2 = -1$. The Chern numbers are numerically calculated by integrating the Berry curvature over the $k_x - k_z$ plane, specifically, $C = \frac{1}{2\pi} \iint dk_x dk_z \, \hat{\mathbf{y}} \cdot (\nabla_{\mathbf{k}} \times \mathcal{A}_{\mathbf{k}})$ where $\mathcal{A}_{\mathbf{k}}$ is the Berry potential [19]. The calculation of the topological numbers in an electromagnetic continuum requires imposing a wave vector cut-off in the nonreciprocal part of the electromagnetic response [19, 20]. Note that the low-frequency band ($\omega = 0^+$) with $C_1 = 1$ is a flat (dark) mode in the *xoz* plane. However, if the propagation direction is slightly tilted so that it gains a small *y*-component the dark mode becomes bright, as can be inferred from the band structure in Fig. (2a*ii*) [52].

Generally, the wave polarization depends on $0º < \chi < 90º$ in a complicated way (it also varies from one band to another). For $\chi = 0º$ (propagation along the *y*-axis parallel to the bias field) the wave is transverse electromagnetic (TEM) and circularly polarized (CP). As may be inferred from Figs. (2a*i*), (2a*ii*) and (2a*iii*), in the range $\omega_{gL} < \omega < \omega_{gU}$ there is a single propagating plane wave in the bulk magnetized plasma for every direction of space. The isofrequency contour associated with this band is shown in Fig. (2b) for the frequency $\omega = (\omega_{gL} + \omega_{gU})/2 = 1.20\omega_p$. Evidently, the associated isofrequency surface has revolution symmetry around the *y*-axis. As seen in Fig. (2b), the isofrequency surface is slightly elongated along the *y*-direction due to the anisotropy of the material. For propagation in the *x*-direction the wave is a linearly polarized (LP) TE-mode, whereas for propagation along the *y*-direction the wave is CP polarized.

### B. *Topological waveguide modes*

Let us now consider the waveguide of Fig. 1a formed by introducing two metallic (PEC) walls at the planes $z = 0$ and $z = d$. The gyrotropic waveguide modes are characterized in Appendix B. Using the developed theory [Eq. (B5)], we determined the dispersion of the waveguide modes in the frequency region $\omega_{gL} < \omega < \omega_{gU}$. A few representative isofrequency contours are shown in Fig. 3a for the case in which the waveguide height is $d = 0.5c/\omega_p$. It turns out that for $\omega_{gL} < \omega < \omega_{gU}$ there is a single propagating mode for every direction in the *xoy* plane. Remarkably, notwithstanding the strong anisotropy of the gyrotropic material, the isofrequency contours are nearly circular, and the anisotropy effects are only noticeable near $\omega \approx \omega_{gL}$ (black contour) wherein the mode enters into cut-off.

Figure 3b shows the detailed mode dispersion in the interval $\omega_{gL} < \omega < \omega_{gU}$ (delimited by the dashed gray lines) for propagation along the *x*- (blue lines) and *y*-direction (dashed green lines). In agreement with the isofrequency contours the two cases are nearly indistinguishable. The numerical simulations show that independent of the propagation direction the dominant component of the electric field is along the *z*-direction, consistent with the boundary conditions imposed by the metallic plates. This orientation of the electric field contrasts sharply with the electric field polarization curves in the bulk gyrotropic material (Fig. 2b). A density plot of a time snapshot of $E_z$ is shown in Fig. 3c for the points *A, B, C* and *D* associated with the isofrequency contour $\omega = 1.2\omega_p$ (see Fig. 3a). The point *B* corresponds to the direction of propagation $\varphi = 60º$ ($\varphi$ is measured with respect to the +*x*-axis) and the coordinate $\tilde{x}$ is measured along this direction. The plots also depict the Poynting vector lines in the waveguide.

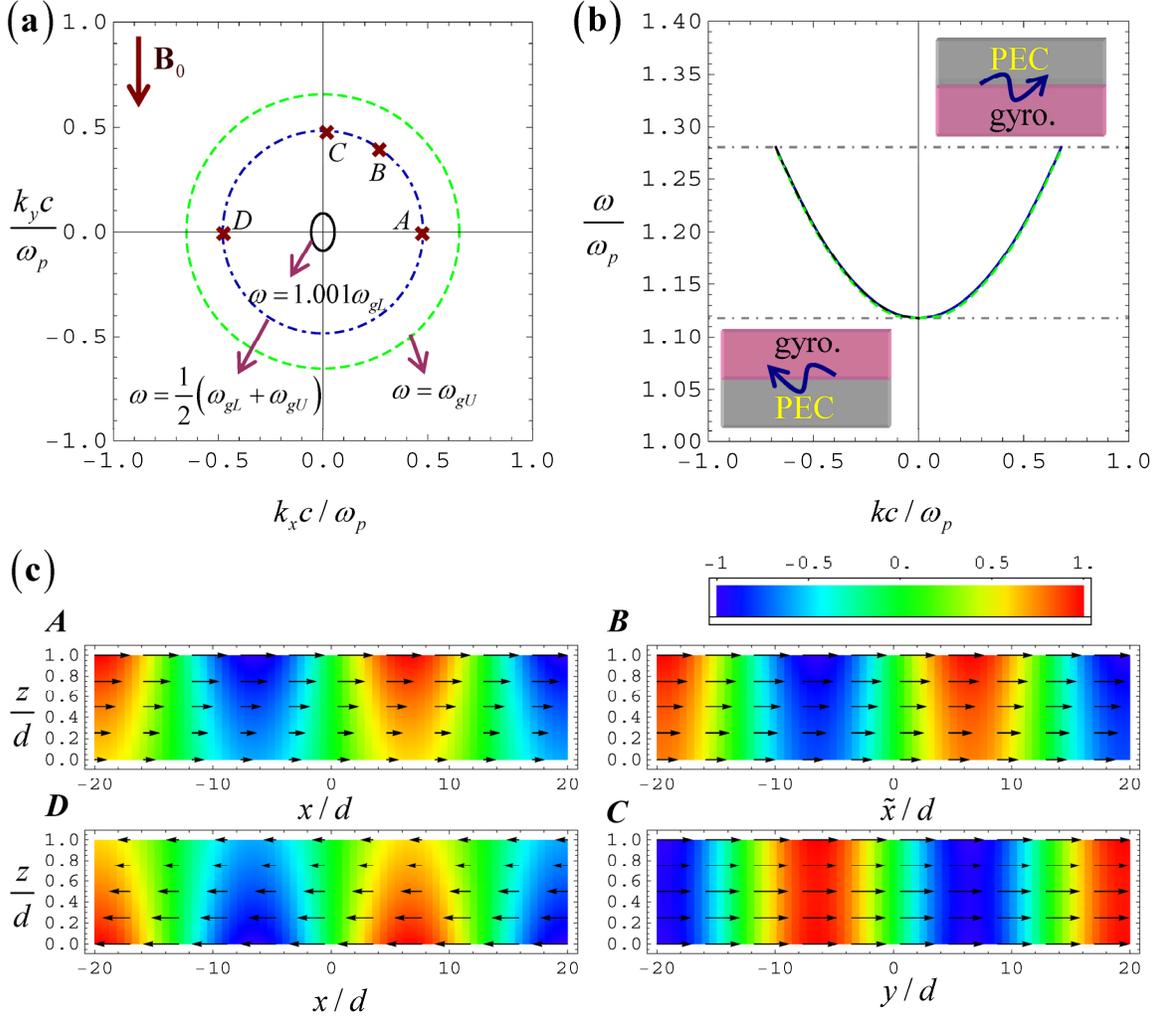

Fig. 3. (Color online) **The gyrotropic waveguide modes.** (a) several isofrequency contours in the range $\omega_{gL} < \omega \leq \omega_{gU}$ for the waveguide height $d = 0.5c/\omega_p$ and for a cyclotron frequency $\omega_0 = 0.5\omega_p$. (b) Mode dispersion for propagation along the *x*-direction (blue lines) and for propagation along the *y*-direction (green dashed lines) [the dispersions in the *x* and *y* directions are nearly coincident]. The black line in the region with a negative abscissa represents the dispersion of the topological edge mode supported when the top plate is removed and is exactly coincident with the corresponding blue line. (c) Time snapshot of the $E_z$ field along the direction of propagation for the modes A, B, C and D marked in panel (a). The arrows represent the Poynting vector lines in the waveguide. The energy tends to be concentrated near the top wall (bottom wall) when the wave propagates along the +*x* (−*x*) direction.

-10-

Notably, the numerical results reveal that for propagation along the +*x* direction (Fig. 3c-A) the energy tends to flow near top wall of the waveguide, whereas for propagation along the –*x* direction (Fig. 3c-D) the bottom wall is preferred. On the other hand, for propagation along *y*-direction the energy distribution is symmetric with respect to the center of the waveguide (Fig. 3c-C). The locking between the spatial region wherein the light flows and its momentum parallels precisely the electron flow induced by a static magnetic field due to incomplete cyclotron orbits (skipping orbits) near the metallic walls (Fig. 1b). This suggests that the two phenomena are deeply linked.

From another perspective, the spatially asymmetric energy transport along the *x*-direction has its origin in the topological properties of the bulk gyrotropic material. Indeed due to the bulk edge correspondence principle [19, 20] a single interface of the gyrotropic material and a standard metal supports exactly one topological edge state in the range $\omega_{gL} < \omega < \omega_{gU}$. Note that the total Chern number below the relevant bandgap is $C_1 + C_2 = -1$, and since a metal is topologically trivial, the Chern number difference is $\delta C = C_{gyro} - C_{metal} = -1$. Thus, a single interface of the two materials must support a unidirectional (TM-polarized) edge mode for propagation along the *x*-direction. The dispersion of the edge mode is determined by [19, 20, 49]:

$$\left( \frac{\gamma_g}{\varepsilon_{ef}} \mp \frac{\varepsilon_{xz} i k_x}{\varepsilon_{xx}^2 + \varepsilon_{xz}^2} \right) + \frac{\gamma_m}{\varepsilon_m} = 0, \qquad (2)$$

where $\varepsilon_m$ is the metal permittivity, $\gamma_m = \sqrt{k_x^2 - \varepsilon_m (\omega/c)^2}$, $\varepsilon_{ef} = (\varepsilon_t^2 - \varepsilon_g^2)/\varepsilon_t$ is the effective permittivity of the gyrotropic material, $\gamma_g = \sqrt{k_x^2 - \varepsilon_{ef}(\omega/c)^2}$, and $\varepsilon_{xx} = \varepsilon_t$ and $\varepsilon_{xz} = i\varepsilon_g$ are the components of the gyrotropic material permittivity tensor. For a perfectly conducting metal the term $\gamma_m / \varepsilon_m$ vanishes. In Eq. (2) the sign $-$ ($+$) is chosen when the metallic wall stands below, in the region $z < 0$ (above, in the region $z > 0$), the gyrotropic material. In particular, the dispersion of the topological edge mode supported by the single-interface waveguide obtained when the top PEC plate is removed is given by $\dfrac{\gamma_g}{\varepsilon_{ef}} + \dfrac{\varepsilon_g k_x}{\varepsilon_{ef} \varepsilon_t} = 0$. This unidirectional topological mode propagates exclusively towards the $-x$ direction and has the dispersion depicted in Fig. 3b with a black line. Notably, this dispersion is precisely coincident (in the region with negative abscissa) with that of the mode supported by the parallel plate gyrotropic waveguide (blue line in Fig. 3b). Indeed, it can be checked that in the limit $k_y \to 0$ the dispersion of the waveguide modes [Eq. (B5)] reduces to $\left(\dfrac{\gamma_g}{\varepsilon_{ef}} + \dfrac{\varepsilon_g}{\varepsilon_{ef} \varepsilon_t} k_x\right)\left(\dfrac{\gamma_g}{\varepsilon_{ef}} - \dfrac{\varepsilon_g}{\varepsilon_{ef} \varepsilon_t} k_x\right) = 0$. Each of the factors can be recognized as the dispersion of the topological state supported by the bottom plate and the gyrotropic material (1st factor) or the top plate and the gyrotropic material (2nd factor). Furthermore, the field distribution of the gyrotropic waveguide modes (for propagation along the *x*-direction) coincides with that of the topological edge states of the individual guides with a single metallic wall. Indeed, when the metal is perfectly conducting the edge state has an electric field oriented along *z*, and as a consequence, the edge state profile is unaffected by the introduction of a second metallic wall. We note in passing that for a metal described by a Drude model the *x* component of the electric field would be nonzero and hence in that scenario the dispersion of the edge modes could be modified by the second metallic plate.

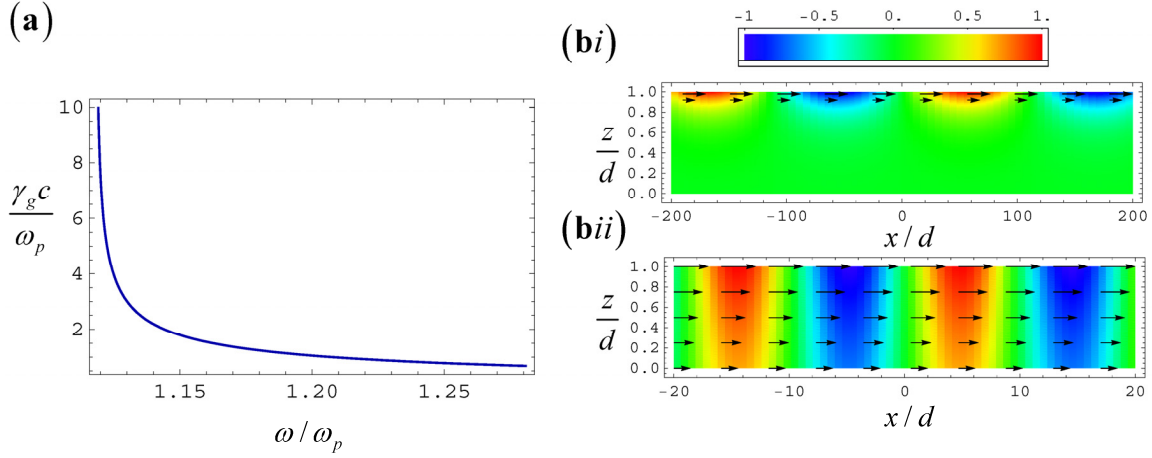

Fig. 4. (Color online) **Mode confinement.** (a) Transverse attenuation constant $\gamma_g$ as a function of the normalized frequency for propagation along the *x*-axis. (b) Time snapshot of $E_z$ for propagation along the +*x*-direction and *(i)* $\omega = 1.001\omega_{gL}$ and *(ii)* $\omega = \omega_{gU}$. The arrows represent the Poynting vector lines in the waveguide.

For propagation along the *x*-direction the confinement (along *z*) of the edge modes is determined by the transverse attenuation constant $\gamma_g$ depicted in Fig. 4a. Somewhat counterintuitively, it is seen that $\gamma_g$ diverges to infinity as the frequency approaches the bottom edge of the TM-waves bandgap ($\omega = \omega_{gL} = 1.12\omega_p$). Note that in the limit $\omega \to \omega_{gL}$ the propagation constant of the edge modes approaches $k_x \to 0$ (Fig. 3b). Thus, different from surface plasmon polaritons, the edge waves become more confined when the guided wavelength is longer. This exotic effect is a consequence of gyrotropic response. Figure 4b shows the modal field profile and the Poynting vector lines near the bandgap edges, showing a dramatic confinement when $\omega \to \omega_{gL}$ (panel *bi*) and weak confinement when $\omega \to \omega_{gU}$ (panel *bii*).

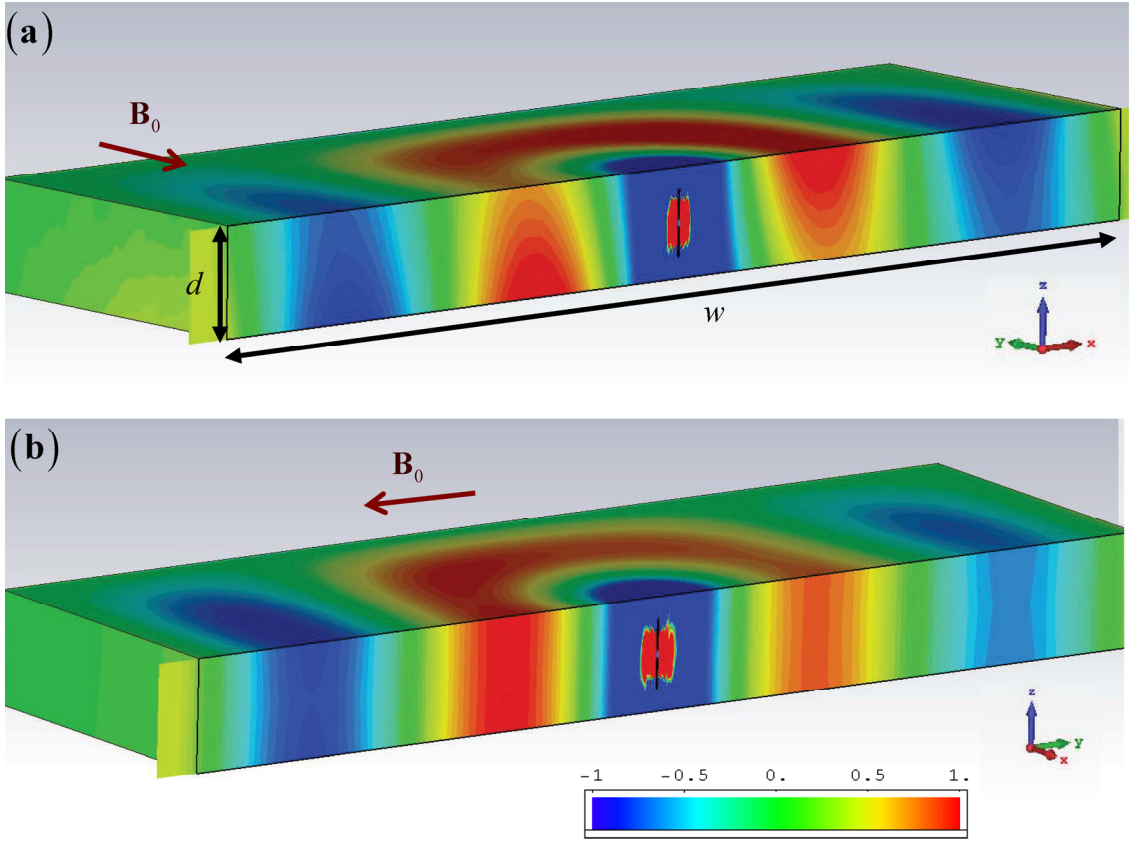

Fig. 5. (Color online) **Field radiated by a vertical electric dipole in the gyrotropic waveguide.** (a) cut of the *xoz* plane. (*b*) cut of the *yoz* plane. The density plots represent a time snapshot of the emitted $E_z$. The waveguide height is $d = 0.5c/\omega_p$ and the cyclotron frequency is $\omega_0 = 0.5\omega_p$. The lateral width of the guide in the plots is $w = 63d$. The oscillation frequency is $\omega = 1.2\omega_p$.

To further illustrate the discussion, we used a commercial electromagnetic simulator [53] to compute the fields radiated by a short vertical electric dipole inside the gyrotropic waveguide. A density plot of a time snapshot of the *z*-component of the radiated electric field is depicted in Fig. 5. Consistent with Fig. 3c, it is seen that in the *xoz* plane (Fig. 5a) the emission in spatially asymmetric, such that the emitted fields flow nearer the top/bottom metallic plates depending if the wave propagates along +*x* or –*x*, respectively. In contrast, in the *yoz* plane the field distribution exhibits an even symmetry with respect to the mid-plane $z = d/2$. The detailed time variation of the emitted field can be found in the time animations available as supplemental materials [54].

In summary, for propagation along the *x*-direction (perpendicular to the bias magnetic field) the waveguide modes are inherently topological. The topological states determine two spatially separated propagation channels, such that the wave energy transported to the positive (negative) *x*-direction is concentrated near the top (bottom) waveguide wall.

## III. Fluctuational Electrodynamics

Having discussed in detail the topological origin of the wave phenomena in the gyrotropic waveguide, next we investigate correlations of the electromagnetic fields induced by either thermal or quantum fluctuations. It is supposed that all the points of the waveguide are held at a fixed temperature *T*.

### *A. Field Correlations*

To begin with, it is convenient to introduce some notations and write the Maxwell's equations in the frequency domain in the following compact form:

$$\hat{N} \cdot \mathbf{F} = \omega \mathbf{M} \cdot \mathbf{F}, \qquad \text{with} \qquad \hat{N} = \begin{pmatrix} 0 & i\nabla \times \mathbf{1}_{3\times 3} \\ -i\nabla \times \mathbf{1}_{3\times 3} & 0 \end{pmatrix}. \tag{3}$$

Here, $\mathbf{1}_{3\times 3}$ is the 3×3 identity matrix, $\omega$ is the oscillation frequency, $\mathbf{F} = \begin{pmatrix} \mathbf{E} & \mathbf{H} \end{pmatrix}^T$ is a six-component vector field written in terms of the standard electromagnetic field vectors, and *T* denotes the transpose operator. The material matrix $\mathbf{M}$ for the geometry of interest is of the form $\mathbf{M}(\mathbf{r},\omega) = \begin{pmatrix} \varepsilon_0 \overline{\varepsilon} & 0 \\ 0 & \mu_0 \mathbf{1}_{3\times 3} \end{pmatrix}$, where $\overline{\varepsilon}(\mathbf{r},\omega)$ is the space dependent permittivity tensor.

In the limit of negligible material loss, the quantized electromagnetic fields in the gyrotropic waveguide may be expressed in terms of the electromagnetic modes $\mathbf{F}_{n\mathbf{k}} = \mathbf{F}_{n\mathbf{k}}(\mathbf{r})$ as [55-58]:

$$\hat{\mathbf{F}}(\mathbf{r},t) = \sum_{\omega_{n\mathbf{k}}>0} \sqrt{\frac{\hbar\omega_{n\mathbf{k}}}{2}} \left(\hat{a}_{n\mathbf{k}} e^{-i\omega_{n\mathbf{k}}t} \mathbf{F}_{n\mathbf{k}}(\mathbf{r}) + \hat{a}^{\dagger}_{n\mathbf{k}} e^{+i\omega_{n\mathbf{k}}t} \mathbf{F}^{*}_{n\mathbf{k}}(\mathbf{r})\right). \tag{4}$$

Here, $\hat{a}_{n\mathbf{k}}, \hat{a}^{\dagger}_{n\mathbf{k}}$ are standard annihilation and creation operators ($\left[\hat{a}_{n\mathbf{k}}, \hat{a}^{\dagger}_{n\mathbf{k}}\right] = 1$) and $\omega_{n\mathbf{k}}$ is the oscillation frequency associated with the mode $n\mathbf{k}$. As previously mentioned, the waveguide modes $\mathbf{F}_{n\mathbf{k}}$ are determined with the formalism of Appendix B. For dispersive media the electromagnetic modes must be normalized as [57, 58]:

$$\frac{1}{2}\int d^3\mathbf{r}\, \mathbf{F}^{*}_{n\mathbf{k}} \cdot \frac{\partial(\omega\mathbf{M})}{\partial\omega} \cdot \mathbf{F}_{n\mathbf{k}} = 1. \tag{5}$$

The above condition ensures that the energy stored in the mode $\mathbf{F}_{n\mathbf{k}} e^{-i\omega_{n\mathbf{k}}t}$ is identical to the unity [19]. Because the waveguide is invariant to translations along $x$ and $y$ the modes can be labeled by the transverse wave vector $\mathbf{k} = (k_x, k_y, 0)$, which determines the fields variation in the $xoy$ plane: $\mathbf{F}_{n\mathbf{k}}(\mathbf{r}) = \mathbf{f}_{n\mathbf{k}}(z) e^{i\mathbf{k}\cdot\mathbf{r}}$. Note that for a cavity with lateral widths ($L_x, L_y$) finite, $k_x, k_y$ are quantized due to the periodic boundary conditions.

Next, we obtain the spectral density of the field correlations in terms of a modal expansion. For simplicity, first it is assumed that the field is in its parametric ground state $|0\rangle$. The ground state depends on the biasing magnetic field. Calculating the Fourier transform of the quantized fields one obtains that $\hat{\mathbf{F}}(\mathbf{r},\omega) = 2\pi \sum_{\omega_{n\mathbf{k}}>0} \sqrt{\frac{\hbar\omega_{n\mathbf{k}}}{2}} \left(\delta(\omega-\omega_{n\mathbf{k}})\hat{a}_{n\mathbf{k}} \mathbf{F}_{n\mathbf{k}} + \delta(\omega+\omega_{n\mathbf{k}})\hat{a}^{\dagger}_{n\mathbf{k}} \mathbf{F}^{*}_{n\mathbf{k}}\right)$. For two generic scalar operators we define the symmetrized product as $\{\hat{A}\hat{B}\} = \frac{1}{2}(\hat{A}\hat{B} + \hat{B}\hat{A})$. Straightforward calculations show that the quantum vacuum expectation ($\langle\ \rangle_0$) of the tensor operator $\{\hat{\mathbf{F}}(\mathbf{r},\omega)\hat{\mathbf{F}}^\dagger(\mathbf{r}',\omega')\}$ is

$$\frac{1}{(2\pi)^2}\left\langle\{\hat{\mathbf{F}}(\mathbf{r},\omega)\hat{\mathbf{F}}^\dagger(\mathbf{r}',\omega')\}\right\rangle_0$$
$$= \delta(\omega-\omega')\mathcal{E}_{0,\omega}\sum_{\omega_{n\mathbf{k}}>0}\frac{1}{2}\left[\delta(\omega-\omega_{n\mathbf{k}})\mathbf{F}_{n\mathbf{k}}(\mathbf{r})\otimes\mathbf{F}^*_{n\mathbf{k}}(\mathbf{r}') + \delta(\omega+\omega_{n\mathbf{k}})\mathbf{F}^*_{n\mathbf{k}}(\mathbf{r})\otimes\mathbf{F}_{n\mathbf{k}}(\mathbf{r}')\right]$$, (6)

where $\mathcal{E}_{0,\omega} = \hbar|\omega|/2$ is the zero-point energy of an harmonic oscillator. In Appendix C it is proven that this result is fully consistent with the fluctuation-dissipation theorem [59], and that the field correlations may also be written in terms of a retarded Green-function. Furthermore, such a link shows that Eq. (6) can be extended to the case of thermally induced fluctuations simply by replacing $\mathcal{E}_{0,\omega}$ by $\mathcal{E}_{T,\omega} = \frac{\hbar\omega}{2}\coth\left(\frac{\hbar\omega}{2k_BT}\right)$ [59], i.e., by the energy of a harmonic oscillator at temperature $T$.

To sum up, in the limit of vanishingly small material losses the field fluctuations can be conveniently characterized using a modal expansion as shown in Eq. (6). This formalism is exactly equivalent to the result obtained using the fluctuation-dissipation theorem. Note that the fluctuation-dissipation theorem [Eq. (C3)] is more general than Eq. (6) because it applies also to systems with strong material absorption.

Calculating the inverse Fourier transforms of the two operators in Eq. (6) it is found that in the time domain:

$$\left\langle\{\hat{\mathbf{F}}(\mathbf{r},t)\hat{\mathbf{F}}(\mathbf{r}',t)\}\right\rangle_T = \int_0^{+\infty}d\omega\,\mathcal{E}_{T,\omega}\sum_{\omega_{n\mathbf{k}}>0}\delta(\omega-\omega_{n\mathbf{k}})\frac{1}{2}\left[\mathbf{F}_{n\mathbf{k}}(\mathbf{r})\otimes\mathbf{F}^*_{n\mathbf{k}}(\mathbf{r}') + \mathbf{F}^*_{n\mathbf{k}}(\mathbf{r})\otimes\mathbf{F}_{n\mathbf{k}}(\mathbf{r}')\right].$$

(7)

The subscript $T$ indicates that the expectation is taken at the temperature $T$, and in accordance, $\varepsilon_{0,\omega}$ was replaced by $\varepsilon_{T,\omega}$.

## *B. Poynting vector*

Using Eq. (7) and $\hat{\mathbf{F}} = \begin{pmatrix} \hat{\mathbf{E}} & \hat{\mathbf{H}} \end{pmatrix}^T$ it is possible to determine the expectation of the Poynting vector operator $\hat{\mathbf{S}} = \{\hat{\mathbf{E}} \times \hat{\mathbf{H}}\}$. It is found that $\langle \hat{\mathbf{S}}(\mathbf{r},t) \rangle_T = \int_0^{+\infty} d\omega\, \mathbf{S}_\omega^+(\mathbf{r})$, being $\mathbf{S}_\omega^+$ the unilateral spectral density of the Poynting vector:

$$\mathbf{S}_\omega^+(\mathbf{r}) = \varepsilon_{T,\omega} \sum_{\omega_{n\mathbf{k}} > 0} \delta(\omega - \omega_{n\mathbf{k}}) \mathbf{S}_{n\mathbf{k}}(\mathbf{r}), \quad \text{with} \quad \mathbf{S}_{n\mathbf{k}}(\mathbf{r}) = \text{Re}\{\mathbf{E}_{n\mathbf{k}}(\mathbf{r}) \times \mathbf{H}_{n\mathbf{k}}^*(\mathbf{r})\}. \tag{8}$$

Here, $\mathbf{E}_{n\mathbf{k}}, \mathbf{H}_{n\mathbf{k}}$ are the electric and magnetic components of the mode $n\mathbf{k}$, i.e., $\mathbf{F}_{n\mathbf{k}} = \begin{pmatrix} \mathbf{E}_{n\mathbf{k}} & \mathbf{H}_{n\mathbf{k}} \end{pmatrix}^T$. In particular, it is seen that the expectation of the Poynting vector operator can be expressed as $\langle \hat{\mathbf{S}}(\mathbf{r},t) \rangle_T = \sum_{\omega_{n\mathbf{k}} > 0} \varepsilon_{T,\omega_{n\mathbf{k}}} \mathbf{S}_{n\mathbf{k}}(\mathbf{r})$. This a physically intuitive result: $\mathbf{S}_{n\mathbf{k}}(\mathbf{r})$ determines the energy density flux when the mode $n\mathbf{k}$ has unit energy [Eq. (5)] and hence $\varepsilon_{T,\omega_{n\mathbf{k}}} \mathbf{S}_{n\mathbf{k}}(\mathbf{r})$ gives the energy density flux when the mode energy is determined by the Bose-Einstein statistics.

Let us first consider that the waveguide is reciprocal (biasing magnetic field is zero), or equivalently (in the limit of vanishing material loss) time-reversal invariant. The time-reversal operator $\mathcal{T}$ transforms the modal fields as $\begin{pmatrix} \mathbf{E}_{n\mathbf{k}} & \mathbf{H}_{n\mathbf{k}} \end{pmatrix} \to \begin{pmatrix} \mathbf{E}_{n\mathbf{k}}^* & -\mathbf{H}_{n\mathbf{k}}^* \end{pmatrix}$ and the frequency and the wave vector as $(\omega_{n\mathbf{k}}, \mathbf{k}) \to (\omega_{n\mathbf{k}}, -\mathbf{k})$. In particular, the Poynting vector is flipped by the time-reversal operation $\mathbf{S}_{n\mathbf{k}} \to -\mathbf{S}_{n\mathbf{k}}$ [17]. Thus, the contribution of the modes $\mathbf{F}_{n\mathbf{k}}$ and $\mathcal{T} \cdot \mathbf{F}_{n\mathbf{k}}$ to the Poynting vector expectation cancels out. Hence, in reciprocal structures the expectation of the Poynting vector is precisely zero at all points of space: $\left\langle \hat{\mathbf{S}}(\mathbf{r},t) \right\rangle_T = 0$ [28]. It can be shown – using directly the fluctuation-dissipation theorem [Eq. (C3)] – that this result remains valid when the material absorption is non-negligible (i.e. when reciprocity is not equivalent to time reversal). Therefore, the system reciprocity needs to be broken to have a radiative heat flow in a stationary state (with the system temperature constant at all points of space) [28].

Since $\mathbf{S}_{n\mathbf{k}}$ determines the Poynting vector of a solution of Maxwell's equations in the frequency domain, it follows that in the absence of material absorption $\nabla \cdot \mathbf{S}_{n\mathbf{k}} = 0$. Thus, from $\left\langle \hat{\mathbf{S}}(\mathbf{r},t) \right\rangle_T = \sum_{\omega_{n\mathbf{k}}>0} \mathcal{E}_{T,\omega_{n\mathbf{k}}} \mathbf{S}_{n\mathbf{k}}(\mathbf{r})$ one concludes that:

$$\nabla \cdot \left\langle \hat{\mathbf{S}}(\mathbf{r},t) \right\rangle_T = 0. \tag{9}$$

Thus, the Poynting vector orbits are necessarily closed, i.e. there are no sources or sinks of the Poynting vector lines. Since the energy flows in closed orbits a nonzero expectation of the Poynting vector does not require pumping energy into the system. The result $\nabla \cdot \left\langle \hat{\mathbf{S}}(\mathbf{r},t) \right\rangle_T = 0$ holds even in the presence of strong material absorption. In that case the rate of energy absorption by a given material element must be precisely the same as the rate of energy emission by the same element to ensure that all material elements are held at a constant temperature.

To further characterize the spectral density of the Poynting vector in the gyrotropic waveguide, it is now supposed that the waveguide widths along the $x$ and $y$ directions are rather large, so that the summation over $\mathbf{k}$ can be transformed into an integral using

$$\sum_{\mathbf{k}} = \frac{A_0}{(2\pi)^2} \int d^2\mathbf{k}$$ where $A_0 = L_x \times L_y$ is the transverse area of the waveguide. Hence, it follows that:

$$\begin{aligned} \mathbf{S}_\omega^+(\mathbf{r}) &= \varepsilon_{T,\omega} \sum_n \frac{1}{(2\pi)^2} \int d^2\mathbf{k}\, \delta(\omega - \omega_{n\mathbf{k}}) A_0 \mathbf{S}_{n\mathbf{k}}(\mathbf{r}) \\ &= \varepsilon_{T,\omega} \frac{1}{d} \sum_n \frac{1}{(2\pi)^2} \int_{\omega = \omega_{n\mathbf{k}}} dl\, \frac{c}{|\nabla_\mathbf{k} \omega_{n\mathbf{k}}|} \tilde{\mathbf{S}}_{n\mathbf{k}}(\mathbf{r}) \end{aligned} \quad (10)$$

The rightmost integral is a line integral (in the wave vector space) over the isofrequency lines $\omega = \omega_{n\mathbf{k}}$. For convenience we introduced the dimensionless vector $\tilde{\mathbf{S}}_{n\mathbf{k}} = \frac{1}{c} V \mathbf{S}_{n\mathbf{k}}$ with $V = A_0 \times d$ the volume of the waveguide. The dimensionless vector $\tilde{\mathbf{S}}_{n\mathbf{k}}$ may be written as $\tilde{\mathbf{S}}_{n\mathbf{k}} = \eta_0 \operatorname{Re}\{\tilde{\mathbf{E}}_{n\mathbf{k}} \times \tilde{\mathbf{H}}_{n\mathbf{k}}^*\}$ ($\eta_0$ is the vacuum impedance) with the modal fields $\tilde{\mathbf{F}}_{n\mathbf{k}} = (\tilde{\mathbf{E}}_{n\mathbf{k}} \ \ \tilde{\mathbf{H}}_{n\mathbf{k}})^T$ normalized as $\frac{1}{\varepsilon_0 V} \frac{1}{2} \int d^3\mathbf{r}\, \tilde{\mathbf{F}}_{n\mathbf{k}}^* \cdot \frac{\partial(\omega \mathbf{M})}{\partial \omega} \cdot \tilde{\mathbf{F}}_{n\mathbf{k}} = 1$. It is evident that when the waveguide is invariant to translations along $x$ and $y$ $\tilde{\mathbf{S}}_{n\mathbf{k}}$ depends exclusively on the $z$ coordinate, so that $\mathbf{S}_\omega^+ = \mathbf{S}_\omega^+(z)$.

## IV. Heat Flow in Closed Orbits

To illustrate the implications of the theory, we used Eq. (10) to determine the Poynting vector spectral density. The calculation is restricted to the frequency interval $\omega_{gL} < \omega < \omega_{gU}$ due to the inherent simplicity of the waveguide dispersion in this frequency range (the waveguide supports precisely one mode per propagation direction) and because of the topological origin of the waveguide modes. The gyrotropic waveguide is characterized by the same parameters as in Sect. II ($\omega_0 = 0.5\omega_p$ and $d = 0.5c/\omega_p$). As expected, the numerical calculations reveal that the Poynting vector expectation has only an x-component: $\mathbf{S}_\omega^+ = S_\omega^+(z)\hat{\mathbf{x}}$. For convenience, we normalize the spectral density $S_\omega^+$ to the temperature dependent density $S_\omega^T = \mathcal{E}_{T,\omega}(\omega_p/c)^2$. Note that the units of $S_\omega^+$ are J/m$^2$.

Equation (10) may be rewritten as $S_\omega^+(z) = \int d\varphi\, \dot{S}_{\omega,\varphi}^+(z)$ where the $\dot{S}_{\omega,\varphi_0}^+(z)d\varphi$ gives the contribution of all the waveguide modes with wave vector $\mathbf{k}$ in the angular sector $\varphi_0 < \varphi < \varphi_0 + d\varphi$. Here, the angle $\varphi$ is measured with respect to the x-direction in the xoy plane. Figure 6a depicts $\dot{S}_{\omega,\varphi}^+(z)$ as a function of $\varphi$ for the case $\omega = 1.2\omega_p$. Consistent with the results of Sect. II, the modes that contribute mostly to the expectation of the Poynting vector near the top wall are associated with $\varphi = 0°$, i.e., are topological waves that propagate along the +x-direction (see the black curve in Fig. 6a). Conversely, near the bottom wall the modes that contribute mostly are associated with $\varphi = 180°$, i.e., with propagation along the –x-direction (dot-dashed blue curve in Fig. 6a). In agreement with the symmetry of the problem, Fig. 6a shows that $\dot{S}_{\omega,\varphi}^+(d) = -\dot{S}_{\omega,\varphi+\pi}^+(0)$. Thus, the fluctuation induced Poynting vector satisfies $S_\omega^+(z) = -S_\omega^+(d-z)$, i.e., has odd symmetry with respect to the center of the waveguide. This property is nicely illustrated by Fig. 6b which shows a density plot of the intensity and the vector lines of $\mathbf{S}_\omega^+$ at $\omega = 1.2\omega_p$. Notably, due to the topological properties of the gyrotropic material the expectation of the spectral density of the Poynting vector is nonzero, and hence there is a net flow of electromagnetic energy in the waveguide upper and lower regions even though the system is in a thermal equilibrium.

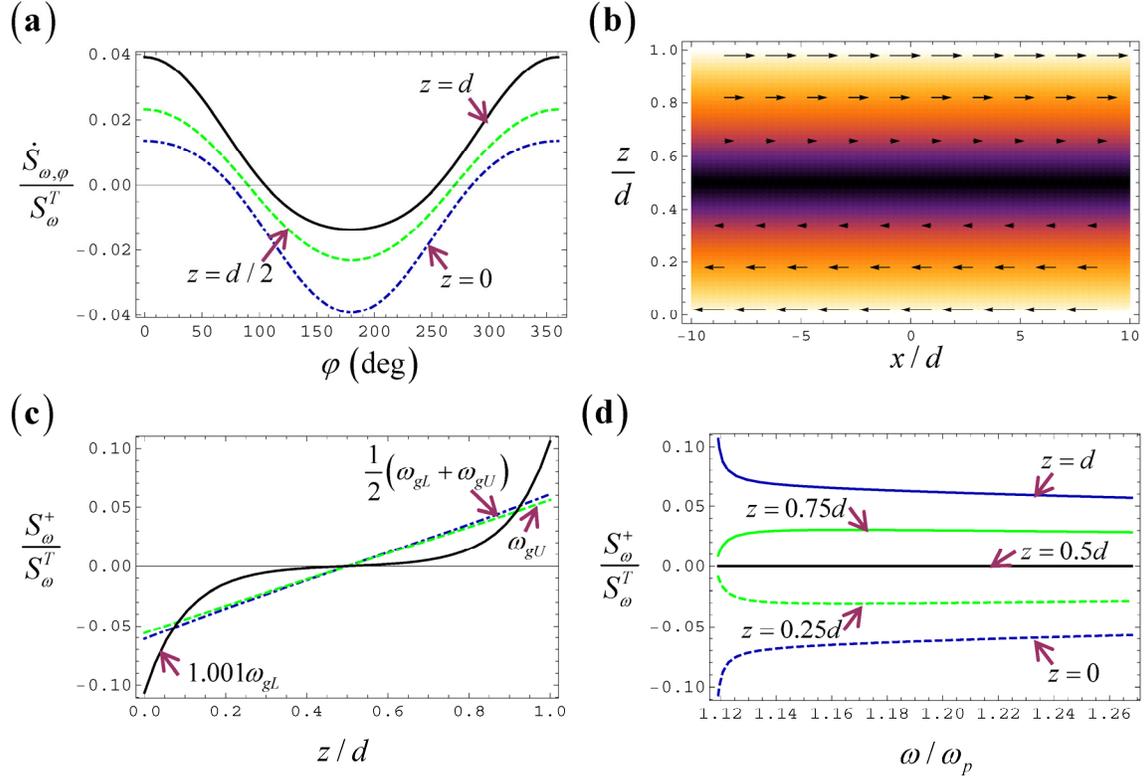

Fig. 6. (Color online) **The fluctuation induced energy density flux.** (a) $\dot{S}^+_{\omega,\varphi}(z)$ as a function of the direction ($\varphi$) of the wave vector for $\omega = 1.2\omega_p$. (b) Density plot of the intensity and vector lines of $\mathbf{S}^+_\omega$ for $\omega = 1.2\omega_p$. (c) $S^+_\omega$ as function of $z$ for $\omega = \omega_{gL}$ (black line), $\omega = 1.2\omega_p$ (dot-dashed blue line) and $\omega = \omega_{gU}$ (dashed green line). (d) $S^+_\omega$ as function of frequency calculated at the positions $z = (m/4)d$ with $m=0,1,\ldots 4$.

As seen in Fig. 6b, the Poynting vector lines form closed orbits oriented in the clockwise direction (the direction opposite to the applied magnetic field). Because the waveguide width is assumed very large along the *x*-direction, the orbits of the Poynting vector are closed at infinity. Remarkably, the heat flow near the top plate is towards the +*x*-direction, imitating thus the electron flow due to incomplete cyclotron orbits near the same metallic wall (see Fig. 1b). Because the electromagnetic momentum density is $\mathbf{S}/c^2$, it follows that even though the total electromagnetic momentum of the system vanishes ($\frac{1}{c^2}\int \mathbf{S} dV = 0$), there is a nontrivial *electromagnetic angular momentum* due to the circulation of energy in closed orbits. The angular momentum is $\mathcal{L} = \frac{1}{c^2}\int \mathbf{r} \times \mathbf{S} dV = \mathcal{L}_y \hat{\mathbf{y}}$ with $\mathcal{L}_y / A_0 = \frac{1}{c^2}\int \left( z - \frac{d}{2} \right) S_x dz$ [60-62] (in this calculation we do not include the contributions at $x = \pm \infty$ from the z-component, $S_z/c^2$, of the momentum density). The system has this attribute even in the limit of a zero temperature, wherein $\mathcal{E}_{T=0,\omega} = \hbar|\omega|/2$ and the fluctuations have a purely quantum origin.

A quantitative characterization of $S_\omega^+$ in the waveguide cross-section is given in Fig. 6c for different frequencies. For $\omega = 1/2(\omega_{gL} + \omega_{gU})$ and $\omega = \omega_{gU}$, $S_\omega^+$ varies approximately linearly with z. In contrast, near the lower bandgap edge $\omega = 1.001\omega_{gL}$ the Poynting vector is strongly depleted near the waveguide center. This property is a consequence of the enhanced confinement of the topological edge modes near $\omega = \omega_{gL}$ (see Fig. 4b).

Figure 6d depicts $S_\omega^+$ as a function of frequency in the band $1.001\omega_{gL} < \omega < \omega_{gU}$ for different values of z. As expected, for $z > d/2$ ($z < d/2$) the spectral density of $S_x$ is positive (negative). Interestingly, in the considered interval $S_\omega^+$ varies slowly with the frequency, except near the $\omega = \omega_{gL}$ wherein, as previously discussed, the guided modes confinement changes dramatically leading to a depletion of the Poynting vector lines near the waveguide center and an enhancement near the metallic plates.

It is relevant to estimate the amount of power that flows in the upper waveguide region. To this end, we may use the approximation $S_\omega^+ \approx 0.06 S_\omega^T (2z/d - 1)$, which is quite satisfactory away from $\omega = \omega_{gL}$. Then, ignoring contributions outside the spectral region $\omega_{gL} < \omega < \omega_{gU}$, it is found that the peak value of the Poynting vector is $dS_x = S_{\omega=1.2\omega_p}^+ \big|_{z=d} d\omega \approx 0.06 \varepsilon_{T,1.2\omega_p} (\omega_p/c)^2 (\omega_{gU} - \omega_{gL})$. For example, if $\omega_p/2\pi = 0.1\,\mathrm{THz}$ and $T = 300\,\mathrm{K}$ one gets $dS_x \approx 0.1\,\mathrm{mW/m^2}$. The magnetic field required to have $\omega_0 = 0.5\omega_p$ is on the order of $B_0 = 1.8\,\mathrm{T}$. Notably, in the limit of a zero temperature, the expectation of the Poynting vector may be still on the order of $dS_x \approx 1\,\mu\mathrm{W/m^2}$. Thus, it seems that the described effect may be within the reach of an experimental verification, perhaps even in the $T \to 0^+$ limit. Generically, the spectral density $S_\omega^+$ in the band gap always increases with $\omega_p$ if the ratio of $\omega_0/\omega_p$ is kept fixed. In practice, the strength of the heat current is limited by the cyclotron frequency and thereby by the static magnetic field, which realistically can hardly be made larger than a few Tesla. A possible experiment may be based on a directional coupler (e.g., created by inserting slots on the top wall of the waveguide [63]) that can be used to detect the imbalance between the heat flows along the +x and –x directions near the top wall.

## V. Discussion and conclusion

In summary, it was demonstrated that in a gyrotropic topological waveguide in thermal equilibrium there is a permanent heat flow in closed orbits. The effect persists even in the limit of a zero temperature wherein the field fluctuations are purely quantum mechanical. This phenomenon is deeply rooted in the nontrivial topological properties of the gyrotropic material, which lead to the formation of unidirectional edge states near each metallic wall. In particular, it was highlighted that in a thermal equilibrium the electromagnetic field has a nontrivial angular momentum, and that the Poynting vector orbits mimic the electron skipping orbits due to an incomplete cyclotron motion. The effect appears to be strong enough to allow for an experimental verification.

We would like to highlight that the considered system is equivalent to a metallic cavity with closed walls in all directions of space, and hence is electromagnetically closed. One may wonder what happens in a system with a single infinitely-wide metallic plate. A conceptual problem with such a configuration is that it is not electromagnetically closed (note also that the gyrotropic material has no full bandgap), and for open systems it is not reasonable to assume that there is a thermal equilibrium to begin with. Nevertheless, the structure can be mathematically closed with periodic boundaries – that impose a cyclic variation of the fields – and which would make the single-plate geometry effectively equivalent to a torus waveguide with a very large radius of curvature. In that case, it is expected that similar to the example studied in the article there is a heat flow in closed orbits (the orbits of the torus) at equilibrium. For a single-plate with finite dimensions along $x$ and $y$ the same conclusion holds, and the heat flux is expected to circulate around the plate in closed orbits (note that in this scenario the topological modes propagating to the left on the top face of the metallic plate are forced to propagate in the opposite direction on the bottom face). Thus, we believe that quantum or thermally induced fluctuations in "one-way" topological systems may lead to exciting new physics, and we hope that this work may stimulate further studies in related directions.

## *Appendix A: Plane wave propagation in a gyrotropic medium*

Consider a generic electric gyrotropic material described by the relative permittivity tensor:

$$\bar{\varepsilon} = \varepsilon_t \mathbf{1}_t + \varepsilon_a \hat{\mathbf{u}}\hat{\mathbf{u}} + i\varepsilon_g \hat{\mathbf{u}} \times \mathbf{1}, \tag{A1}$$

where $\mathbf{1}_t = \mathbf{1} - \hat{\mathbf{u}}\hat{\mathbf{u}}$. The unit vector $\hat{\mathbf{u}}$ determines the direction of the bias static magnetic field. The elements $\varepsilon_t, \varepsilon_a$ determine the transverse and axial (with respect to the bias field) permittivity components, whereas $\varepsilon_g$ determines the gyrotropic response. A plane wave in the gyrotropic medium satisfies:

$$\mathbf{k} \times \mathbf{E} = \omega\mu_0 \mathbf{H}, \qquad \mathbf{k} \times \mathbf{H} = -\omega\varepsilon_0 \bar{\varepsilon} \cdot \mathbf{E} \tag{A2}$$

where $\mathbf{k} = \mathbf{k}_t + k_u \hat{\mathbf{u}}$ is the wave vector, being $k_u = \mathbf{k} \cdot \hat{\mathbf{u}}$ and $\mathbf{k}_t = \hat{\mathbf{u}} \times (\mathbf{k} \times \hat{\mathbf{u}})$ its longitudinal and transverse components. Thus, it follows that $\mathbf{k} \times (\mathbf{k} \times \mathbf{E}) + \frac{\omega^2}{c^2} \bar{\varepsilon} \cdot \mathbf{E} = 0$. Looking for solutions of the form $\mathbf{E} = \alpha_1 \mathbf{k} \times \hat{\mathbf{u}} + \alpha_2 \mathbf{k}_t + \alpha_3 \hat{\mathbf{u}}$ it is found after some algebra that the electric field must be of the form:

$$\mathbf{E} \sim \frac{i\varepsilon_g \frac{\omega^2}{c^2}}{\frac{\omega^2}{c^2}\varepsilon_t - k^2} \mathbf{k} \times \hat{\mathbf{u}} + \mathbf{k}_t + \frac{-k_u k_t^2}{\frac{\omega^2}{c^2}\varepsilon_a - k_t^2} \hat{\mathbf{u}}, \tag{A3}$$

and that the wave vector satisfies the following dispersion equation:

$$\left(\varepsilon_t^2 - \varepsilon_g^2\right)\varepsilon_a \frac{\omega^4}{c^4} - \left[\left(\varepsilon_t\left(\varepsilon_t + \varepsilon_a\right) - \varepsilon_g^2\right)k_t^2 + 2\varepsilon_t \varepsilon_a k_u^2\right]\frac{\omega^2}{c^2} + \left(\varepsilon_t k_t^2 + \varepsilon_a k_u^2\right)\left(k_t^2 + k_u^2\right) = 0. \tag{A4}$$

## Appendix B: Modes of the gyrotropic waveguide

Here, we determine modes in a gyrotropic waveguide with the geometry of Fig. 1. It is supposed that the bias magnetic field is parallel (or anti-parallel) to $\hat{\mathbf{u}} = \hat{\mathbf{y}}$. The waveguide modes depend on $x$ and $y$ as $e^{ik_x x} e^{ik_y y}$ and hence in the region between the metallic plates the fields can be written as a superposition of four plane waves of the bulk gyrotropic medium [see the Appendix A] with wave vector components $k_u = k_y$ and $\mathbf{k}_t = k_x \hat{\mathbf{x}} \pm k_{z,i} \hat{\mathbf{z}}$ ($i$=1, 2). From Eq. (A4) one finds that $k_{z,i}$ is required to satisfy:

$$k_{z,i}^2 = -k_x^2 + \frac{1}{2\varepsilon_t} \left[ \left( \varepsilon_t (\varepsilon_t + \varepsilon_a) - \varepsilon_g^2 \right) \frac{\omega^2}{c^2} - (\varepsilon_a + \varepsilon_t) k_y^2 \right]$$

$$\pm \frac{1}{2\varepsilon_t} \sqrt{\left[ \left( \varepsilon_t (\varepsilon_t + \varepsilon_a) - \varepsilon_g^2 \right) \frac{\omega^2}{c^2} - (\varepsilon_a + \varepsilon_t) k_y^2 \right]^2 - 4\varepsilon_t \left[ (\varepsilon_t^2 - \varepsilon_g^2) \varepsilon_a \frac{\omega^4}{c^4} - 2\varepsilon_t \varepsilon_a k_y^2 \frac{\omega^2}{c^2} + \varepsilon_a k_y^4 \right]}.$$

(B1)

Each of the four possible solutions $\pm k_{z,i}$ ($i$=1, 2) is associated with a plane wave with electric field determined by Eq. (A3). Hence, the electric field in the gyrotropic waveguide is of the form (the variation on $x$ and $y$ is omitted):

$$\mathbf{E} = \left( \Delta_1 \mathbf{k}_1^+ \times \hat{\mathbf{y}} + \mathbf{k}_{t,1}^+ + \theta_1 k_y \hat{\mathbf{y}} \right) A_1^+ e^{ik_{z,1} z} + \left( \Delta_1 \mathbf{k}_1^- \times \hat{\mathbf{y}} + \mathbf{k}_{t,1}^- + \theta_1 k_y \hat{\mathbf{y}} \right) A_1^- e^{-ik_{z,1} z} +$$
$$\left( \Delta_2 \mathbf{k}_2^+ \times \hat{\mathbf{y}} + \mathbf{k}_{t,2}^+ + \theta_2 k_y \hat{\mathbf{y}} \right) A_2^+ e^{ik_{z,2} z} + \left( \Delta_2 \mathbf{k}_2^- \times \hat{\mathbf{y}} + \mathbf{k}_{t,2}^- + \theta_2 k_y \hat{\mathbf{y}} \right) A_2^- e^{-ik_{z,2} z}.$$

(B2)

where $\mathbf{k}_{t,i}^\pm = k_x \hat{\mathbf{x}} \pm k_{z,i} \hat{\mathbf{z}}$ is the transverse wave vector (with respect to the bias field), $\mathbf{k}_i^\pm = \mathbf{k}_{t,i}^\pm + k_y \hat{\mathbf{y}}$ is the wave vector, $A_i^\pm$ ($i$=1,2) are coefficients of the expansion, and the parameters $\theta_i$ and $\Delta_i$ are defined as

$$\Delta_i = \frac{i\varepsilon_g \frac{\omega^2}{c^2}}{\frac{\omega^2}{c^2}\varepsilon_t - (k_y^2 + k_{t,i}^2)}, \qquad \theta_i = \frac{-k_{t,i}^2}{\frac{\omega^2}{c^2}\varepsilon_a - k_{t,i}^2}, \qquad (B3)$$

with $k_{t,i}^2 = k_x^2 + k_{z,i}^2$. Imposing that the tangential components ($E_x, E_y$) of the electric field vanish at the metallic plates ($z = 0$ and $z = d$) it is found that (for $k_y \neq 0$):

$$\begin{pmatrix} -k_{z,1}\Delta_1 + k_x & k_{z,1}\Delta_1 + k_x & -k_{z,2}\Delta_2 + k_x & k_{z,2}\Delta_2 + k_x \\ \left(-k_{z,1}\Delta_1 + k_x\right)e^{ik_{z,1}d} & \left(k_{z,1}\Delta_1 + k_x\right)e^{-ik_{z,1}d} & \left(-k_{z,2}\Delta_2 + k_x\right)e^{ik_{z,2}d} & \left(k_{z,2}\Delta_2 + k_x\right)e^{-ik_{z,2}d} \\ \theta_1 & \theta_1 & \theta_2 & \theta_2 \\ \theta_1 e^{ik_{z,1}d} & \theta_1 e^{-ik_{z,1}d} & \theta_2 e^{ik_{z,2}d} & \theta_2 e^{-ik_{z,2}d} \end{pmatrix} \begin{pmatrix} A_1^+ \\ A_1^- \\ A_2^+ \\ A_2^- \end{pmatrix} = 0$$

(B4)

In particular, by setting the determinant of the matrix equal to zero one obtains the modal dispersion equation:

$$\frac{1}{k_{t,1}^2 k_{t,2}^2}\left[\Delta_1^2 k_{z,1}^2 \theta_2^2 + \Delta_2^2 k_{z,2}^2 \theta_1^2 - k_x^2 \left(\theta_1 - \theta_2\right)^2\right]\frac{\sin(k_{z1}d)}{k_{z1}}\frac{\sin(k_{z2}d)}{k_{z2}}$$
$$+ 2\frac{\theta_1 \theta_2 \Delta_1 \Delta_2}{k_{t,1}^2 k_{t,2}^2}\left(\cos(k_{z1}d)\cos(k_{z2}d) - 1\right) = 0$$

(B5)

For each solution of the modal equation one can find the null space of the matrix in Eq. (B4) and in this manner the electric field. The magnetic field can be obtained from Eq. (B2) taking into account that for each plane wave $\mathbf{H} = \mathbf{k}\times\mathbf{E}/(\omega\mu_0)$.

## *Appendix C: Link with the fluctuation-dissipation theorem*

Here, it is shown that the right-hand side of Eq. [(6)] can be written in terms of a (retarded) Green function $\overline{G}$ that satisfies $\hat{N}\cdot\overline{G} - \omega\mathbf{M}\cdot\overline{G} = \frac{1}{\omega}\mathbf{1}_{6\times 6}\delta(\mathbf{r}-\mathbf{r}')$. To prove this property we use the fact that the electrodynamics of a lossless system is determined by a Hermitian operator, even in presence of material dispersion [19, 64, 65]. Thus, the Green function can be expanded in terms of modes as follows (the details can be found in Appendix B of Ref. [58]):

$$\overline{G}(\mathbf{r},\mathbf{r}') = \frac{1}{2\omega} \sum_{n\mathbf{k}} \frac{1}{\omega_{n\mathbf{k}} - \omega} \mathbf{F}_{n\mathbf{k}}(\mathbf{r}) \otimes \mathbf{F}_{n\mathbf{k}}^*(\mathbf{r}')$$
$$= \frac{1}{2\omega^2} \sum_{\omega_{n\mathbf{k}} \neq 0} \frac{\omega_{n\mathbf{k}}}{\omega_{n\mathbf{k}} - \omega} \mathbf{F}_{n\mathbf{k}}(\mathbf{r}) \otimes \mathbf{F}_{n\mathbf{k}}^*(\mathbf{r}') - \frac{1}{\omega^2} \mathbf{M}_\infty^{-1} \delta(\mathbf{r}-\mathbf{r}')$$
(C1)

The modes $\mathbf{F}_{n\mathbf{k}}$ are normalized as in Eq. (5) and in the first identity the summation includes all modes (positive, negative and zero frequencies). The second identity restricts the summation to modes with nonzero frequencies and is obtained using $\frac{1}{2}\sum_{n\mathbf{k}} \mathbf{F}_{n\mathbf{k}}(\mathbf{r}) \otimes \mathbf{F}_{n\mathbf{k}}^*(\mathbf{r}') = \mathbf{M}_\infty^{-1} \delta(\mathbf{r}-\mathbf{r}')$ (see Ref. [58, Eq. (A8)]). The matrix $\mathbf{M}_\infty$ is defined as $\mathbf{M}_\infty = \lim_{\omega \to \infty} \mathbf{M}(\omega)$. From Eq. (C1), it is simple to check that:

$$\overline{G}(\mathbf{r},\mathbf{r}') - \overline{G}^\dagger(\mathbf{r}',\mathbf{r}) = \frac{i\pi}{\omega} \sum_{\omega_{n\mathbf{k}} \neq 0} \delta(\omega - \omega_{n\mathbf{k}}) \mathbf{F}_{n\mathbf{k}}(\mathbf{r}) \otimes \mathbf{F}_{n\mathbf{k}}^*(\mathbf{r}')$$
$$= \frac{i\pi}{\omega} \sum_{\omega_{n\mathbf{k}} > 0} \left[ \delta(\omega - \omega_{n\mathbf{k}}) \mathbf{F}_{n\mathbf{k}}(\mathbf{r}) \otimes \mathbf{F}_{n\mathbf{k}}^*(\mathbf{r}') + \delta(\omega + \omega_{n\mathbf{k}}) \mathbf{F}_{n\mathbf{k}}^*(\mathbf{r}) \otimes \mathbf{F}_{n\mathbf{k}}(\mathbf{r}') \right]$$
(C2)

The rightmost identity is obtained noting that the reality of the electromagnetic fields implies that negative frequency modes are linked to positive frequency modes by complex conjugation. Thus, it follows from Eq. (6) that:

$$\frac{1}{(2\pi)^2} \left\langle \left\{ \hat{\mathbf{F}}(\mathbf{r},\omega) \hat{\mathbf{F}}^\dagger(\mathbf{r}',\omega') \right\} \right\rangle_0 = \delta(\omega - \omega') \varepsilon_{0,\omega} \frac{\omega}{2\pi i} \left[ \overline{G}(\mathbf{r},\mathbf{r}',\omega) - \overline{G}^\dagger(\mathbf{r}',\mathbf{r},\omega) \right]. \quad \text{(C3)}$$

This equation is fully consistent with the fluctuation-dissipation theorem result [59].

**Acknowledgements:** This work was funded by Fundação para a Ciência e a Tecnologia under project PTDC/EEI-TEL/4543/2014 and by Instituto de Telecomunicações under project UID/EEA/50008/2013.